\documentclass[cameraready]{Interspeech}

\usepackage{amsmath,amssymb}
\usepackage{graphicx}
\usepackage{booktabs}
\usepackage{multirow}
\usepackage{siunitx}
\usepackage{arydshln}
\usepackage{tikz}
\usetikzlibrary{tikzmark}
\usepackage{url}
\usepackage{hyperref}
\usepackage{placeins}
\usepackage{capt-of}

\hypersetup{
  colorlinks=true,
  linkcolor=black,
  citecolor=black,
  urlcolor=black
}

\title{\vspace{-0.6in}SEAM: Shortcut-Aware Real-Time Detection of Scripted vs.\ Spontaneous Speech for Interview Guardrails}
\author[affiliation={1}, correspondingauthor]{Vsevolod}{Kovalev}
\author[affiliation={1,2}]{Pranay}{Manocha}

\address{
  $^1$ Symbal AI, USA\\
  $^2$ Princeton University
}
\newcommand{\ignorethis } [1] {}

\newcommand{\Reals      }     {{\textrm{I\kern-0.18em R}}}

\renewcommand{\vec      } [1] {{\text{\boldmath $\mathbit{#1}$}}}

\newcommand{\change     } [1] {\mbox{{\footnotesize $\Delta$} \kern-3pt}#1}

\makeatletter
\newcommand*{\etc}{%
    \@ifnextchar{.}%
        {etc}%
        {etc.\@\xspace}%
}
\makeatother

\newcommand*{\skippingparagraph}{\par\vspace{0.5\baselineskip}\noindent}

\email{vsevolod.kovalev@outlook.com, pranay@symbal.ai}


\definecolor{seamred}{rgb}{0.55,0.10,0.10}
\definecolor{seamgreen}{rgb}{0.10,0.42,0.12}
\newcommand{\bred}[1]{\textcolor{seamred}{\textbf{#1}}}
\newcommand{\bgreen}[1]{\textcolor{seamgreen}{\textbf{#1}}}

\begin{document}
\ninept
\maketitle

%===============================================================================
% Abstract (<=1000 chars, ASCII only, no citations, must match submission form)
\begin{abstract}

Scripted vs spontaneous speech detection is appealing for interview guardrails, but benchmark performance can be inflated by shortcuts tied to corpus identity, channel conditions, and recording artifacts rather than speaking style itself. We present \textsc{SEAM}, a shortcut-aware framework for real-time scriptedness detection that combines uniform preprocessing, seam-aware sampling, non-speech augmentation, and a compact DistilHuBERT backbone. With 8s windows, the model achieves 0.971 $\pm$ 0.004 ROC-AUC on an external interview-domain evaluation set. Removing the shortcut-prevention components improves internal held-out metrics but sharply reduces external performance, indicating shortcut learning. Post-training quantization reduces the model footprint to 41.8\,MB with little loss in external performance. The results demonstrate that robust real-time scriptedness detection depends not only on the backbone, but on shortcut-aware data design and evaluation. We release code and model checkpoints.

\end{abstract}

\vspace{-0.1in}
\section{Introduction}

Distinguishing scripted speech from spontaneous speech is useful in applications such as corpus curation, speaking-style analysis, and guardrails for conversational systems~\cite{elisha_multilingual,speaking_stype}. In AI-assisted job interviews, this distinction has a particularly practical role: a lightweight real-time audio model can flag stretches of speech that sound read or heavily rehearsed, providing a narrow integrity signal for human review rather than a standalone decision-maker. This use case places the task in a deployment setting where both robustness and low latency matter.

The difficulty, however, is not simply recognizing differences in prosody, fluency, or temporal organization between scripted and spontaneous speech. In real datasets, these labels are often entangled with microphone characteristics, room acoustics, codec effects, corpus genre, speaker population, and other nuisance variables~\cite{prosidic_features,audio_prosidy}. As a result, a model can achieve strong held-out accuracy while relying on spurious shortcuts such as ``clean studio audio'' versus ``noisy conversational audio'' instead of genuine speaking-style cues. For deployment, this is a serious failure mode: benchmark gains may not survive even modest shifts in corpus, channel, or recording conditions.

Recent self-supervised speech encoders have made audio classification substantially stronger~\cite{ssl,distilhubert}, but backbone quality alone does not solve this problem. A larger or better pretrained model can still learn the wrong decision rule when style labels remain confounded with dataset artifacts. For real-time interview guardrails, the challenge is therefore twofold: the system must be compact enough to run alongside other live speech components, and its training and evaluation pipeline must actively discourage shortcut learning. We argue that robust scripted-versus-spontaneous detection depends as much on data construction and evaluation design as on the encoder itself.

We address this problem with \textbf{SEAM}\footnote{Code and checkpoints available at \url{https://github.com/vsevolod-kovalev/seam}.} (\textbf{S}hortcut-aware \textbf{E}valuation, \textbf{A}ugmentation, and \textbf{M}odeling), a framework for real-time scriptedness detection built around coordinated choices in preprocessing, sampling, augmentation, modeling, and evaluation. SEAM combines uniform waveform preprocessing to reduce channel-level variation, provenance-tracked seam-aware sampling to prevent artificial joins from becoming style cues, a dedicated non-speech noise bank to weaken ``clean audio implies scripted speech'' heuristics, and evaluation on both grouped internal splits and an external interview-domain benchmark. These components are paired with a compact DistilHuBERT-based classifier chosen for favorable quality--efficiency tradeoffs in live use.

Our contributions are as follows:
\begin{itemize}
\item We present \textsc{SEAM}, a shortcut-aware pipeline for real-time scripted-versus-spontaneous speech detection that integrates preprocessing, seam-aware sampling, non-speech augmentation, compact modeling, and transfer-oriented evaluation.
\item We show that shortcut-prevention components improve external interview-domain transfer even when they reduce internal held-out performance, demonstrating that naive benchmark gains reflect shortcut learning than better style modeling.
\item We identify a compact DistilHuBERT configuration (23\,M parameters) that achieves strong internal and external performance while remaining practical for low-latency deployment.
\item We provide ablation and deployment evidence, including partial fine-tuning studies and post-training quantization, showing that robust scriptedness detection can be achieved within realistic real-time memory and latency budgets.
\end{itemize}

\vspace{-0.1in}
\section{Related Work}

\subsection{Scripted vs.\ spontaneous speech classification}
Prior work has studied scripted, read, narrated, and spontaneous speech classification as a speaking-style recognition problem motivated by corpus analysis and downstream media applications. The closest recent reference point is Elisha \emph{et al.}~\cite{elisha_multilingual}, who show that transformer-based audio models perform strongly for spontaneous-versus-scripted classification at multilingual scale. Our setting is narrower and more deployment-driven: we focus on English interview audio, real-time operation, and robustness to recording and corpus shortcuts rather than benchmark performance alone.

\vspace{-0.1in}

\subsection{Compact self-supervised speech models}
Self-supervised speech encoders such as wav2vec2.0, HuBERT, and WavLM have become strong front-ends for downstream audio classification~\cite{wav2vec2,hubert,wavlm,ssl,hear,content_representation}. Distillation further improves deployability by reducing model size while preserving much of the downstream utility of larger encoders; DistilHuBERT is a representative example~\cite{distilhubert}. Our work builds on this line, but asks a more specific question: whether a compact SSL model can remain reliable for scripted-versus-spontaneous detection when training and evaluation are designed to suppress shortcut learning.

\vspace{-0.1in}

\subsection{Shortcut learning and robustness in speech/audio}
Speech and audio models are often sensitive to nuisance variables such as corpus identity, speaker population, microphone characteristics, room acoustics, silence structure, and other recording artifacts. Prior work has addressed related robustness issues through augmentation, domain-invariant training, and confound-aware evaluation~\cite{specaugment,dann,yang_bias_2024,polle24_confounds,shim23_shortcuts}. Our paper is closest in spirit to this robustness-oriented perspective: rather than treating preprocessing and data construction as implementation details, we make shortcut-aware preprocessing, seam-aware sampling, non-speech augmentation, and transfer-oriented evaluation central components of the method.

\section{The SEAM Framework}
\label{sec:seam}

We refer to our end-to-end pipeline as \textbf{SEAM}. The core idea is that robust scripted-vs.-spontaneous detection does not come from the backbone alone, but from coordinated choices in task design, data curation, preprocessing, sampling, augmentation, and evaluation. SEAM is designed for real-time interview guardrails, where the model must operate with low latency, remain lightweight enough to run alongside other speech components, and avoid relying on corpus- or channel-specific shortcuts.

SEAM targets three recurring failure modes in this task. First, \emph{channel shortcuts}: microphone characteristics, room acoustics, loudness, codec artifacts, and long non-speech regions can become proxies for the label. Second, \emph{provenance shortcuts}: when training windows cross recording boundaries or inherit dataset-specific structure, the model can learn seams or corpus fingerprints instead of speaking style. Third, \emph{evaluation shortcuts}: standard held-out accuracy can remain high when the model relies on spurious cues that fail under corpus or channel shift. SEAM addresses these with three corresponding defenses: uniform waveform preprocessing to suppress channel variation, provenance-tracked seam-aware sampling with non-speech augmentation to suppress recording and cleanliness heuristics, and transfer-oriented evaluation that explicitly tests whether performance survives beyond the internal benchmark.

\subsection{Framework design and task}
Given an audio segment $x(t)$, the task is to predict a binary label $y \in \{0,1\}$ for scripted $(1)$ versus spontaneous $(0)$ speech. In our training data the scripted class is operationally read-aloud speech (audiobooks and Spoken Wikipedia readings), but the task targets any pre-composed delivery, read or rehearsed, and the external benchmark is annotated on perceived scriptedness spanning both. We treat these labels as coarse speaking-style categories rather than perfectly discrete states: mixed or transitional regions can occur within a file, which motivates window-level scoring followed by aggregation over longer spans. In deployment, inference is performed on fixed-length windows and then aggregated across multiple windows or speaker turns. The resulting score is intended as a narrow guardrail signal for human review rather than a standalone decision-maker. This setting imposes two practical requirements. The model must be \emph{real-time capable}, which favors compact self-supervised encoders and shallow task adaptation, and it must be \emph{shortcut-resistant}, because apparent performance gains that come from corpus or channel cues will not survive target-domain shift.

\subsection{Shortcut-aware data processing}
We curate four English corpora: spontaneous speech from People's Speech (filtered)~\cite{peoples_speech} and PodcastFillers (CC-BY subset)~\cite{podcastfillers}, and scripted speech from LibriSpeech~\cite{librispeech} and Spoken Wikipedia (English)~\cite{swc_lrec,swc_lre}. We restrict training to subsets with licenses permitting commercial reuse and exclude NC/ND subsets when present. To weaken easy channel heuristics, we deliberately pair corpora with partially overlapping acoustic profiles; for example, podcast speech can be relatively clean, while read speech in Spoken Wikipedia can still contain diverse microphones and room conditions.

All audio is converted to mono 16\,kHz and passed through a uniform waveform preprocessing pipeline consisting of (i) DC removal, (ii) a 70\,Hz high-pass biquad filter, (iii) integrated loudness normalization to $-23$\,LUFS for non-silent material, and (iv) peak limiting at 0.99 followed by clamping to $[-1,1]$~\cite{dpam_pm}. Where applicable, dataset-specific intro/outro material and long pauses are removed to reduce non-speech and corpus-specific shortcuts. These steps are intended to weaken gain-, codec-, and room-dependent cues while preserving the prosodic and temporal structure relevant to speaking style.

For scalable training, processed audio is stored as fixed 10-minute FLAC chunks and packed into 20-hour tar shards with JSONL manifests that preserve provenance, including recording ids, offsets, corpus ids, and grouping keys. Training windows are sampled only from provenance-consistent segments and are constrained to remain within a single source recording. This seam-aware sampling is important because spontaneous corpora often contain shorter source recordings than scripted corpora; without such constraints, cross-recording joins and artificial boundaries can become unintended style cues. If a target window cannot be drawn cleanly from a single provenance segment, we do not join across recordings. Instead, the remainder of the window is padded with non-speech material.

% 80(train)-10(eval)-10(test)

To further break the heuristic that ``clean audio implies scripted speech,'' we construct a dedicated non-speech noise bank using a SpeechBrain CRDNN voice activity detector~\cite{speechbrain}. We collect approximately 14 hours of non-voice material, including room tone, breathing, and microphone artifacts, while explicitly excluding speech. During training, a randomly selected noise clip is mixed in additively at a sampled SNR over a random 40--70\% of the window.

% hours + speakers + role per fold

\subsection{Model architecture and training procedure}
SEAM uses a pretrained self-supervised speech encoder followed by temporal mean pooling and a lightweight classifier head:
\begin{equation}
z = \mathrm{meanpool}(\mathrm{Encoder}(x_{1:T}))
\end{equation}
\vspace{-0.1in}
\begin{equation}
\ell_{\mathrm{audio}} = \mathrm{MLP}(z)
\end{equation}
The classifier head is a 2-layer MLP with ReLU activations and dropout ($p=0.30$), trained using BCEWithLogitsLoss. Optimization uses AdamW~\cite{adamw} with weight decay 0.01. We use separate, constant learning rates of $5\times10^{-6}$ for the encoder and $3\times10^{-4}$ for the classification head.

We screen several frozen SSL backbones under a fixed training budget and use WavLM Base+ as a quality reference, but select DistilHuBERT for the main system because it offers the best practical quality-efficiency tradeoff for real-time deployment. To preserve this advantage, we avoid full end-to-end fine-tuning and instead study shallow adaptation regimes: head-only training, head plus the top transformer layer, head plus the top two transformer layers, and a more aggressive variant that additionally exposes part of the convolutional frontend.

The final configuration uses DistilHuBERT with 8\,s windows, unfreezing only the top transformer layer, and enabling the noise-bank augmentation. The reported final model is trained for three epochs across three random seeds on an NVIDIA A100 80GB. Faster ablations and backbone screening use a fixed-budget regime with batch size 256, 4 shards per class (80 hours per class), and 120 optimization steps. We report window-level Accuracy and ROC-AUC; in deployment, these scores are aggregated over multiple windows.

% Training uses AdamW with ...
% We apply ...
% The MLP head uses dropout p=...
% Models are selected by eval AUC.

% \begin{table}[!t]
% \centering
% \footnotesize
% \setlength{\tabcolsep}{4pt}
% \resizebox{\columnwidth}{!}{%
% \begin{tabular}{lcccccc}
% \toprule
% Seed & {\bf Eval Acc} & {\bf Eval AUC} & {\bf Test Acc} & {\bf Test AUC} & {\bf Ext Acc} & {\bf Ext AUC} \\
% \midrule
% 1337 & 0.9690 & 0.9820 & 0.9639 & 0.9781 & 0.9540 & 0.9725 \\
% 1338 & 0.9602 & 0.9762 & 0.9548 & 0.9715 & 0.9431 & 0.9669 \\
% 1339 & 0.9728 & 0.9835 & 0.9681 & 0.9802 & 0.9580 & 0.9745 \\
% \midrule
% \tikzmarknode{meanL}{Mean} & 0.9673 & 0.9806 & 0.9623 & 0.9766 & 0.9517 & \tikzmarknode{meanR}{0.9713} \\
% \noalign{%
% \begin{tikzpicture}[overlay,remember picture]
% \draw[densely dotted, rounded corners, line width=0.8pt]
% ([xshift=-2pt,yshift=2pt]meanL.north west) rectangle
% ([xshift=2pt,yshift=-2pt]meanR.south east);
% \end{tikzpicture}%
% }
% Std & 0.0065 & 0.0039 & 0.0068 & 0.0045 & 0.0077 & 0.0039 \\
% \bottomrule

% \end{tabular}%
% }
% \caption{Performance of the final \textsc{SEAM} model under the full-training regime. The model uses DistilHuBERT with 8\,s windows and top-1-layer unfreezing, trained for 3 epochs. Metrics are reported on the grouped internal \textit{eval}/\textit{test} splits and the external interview-domain set; mean and standard deviation are computed across three seeds.}
% \label{tab:final}
% \vspace{-0.1in}
% \end{table}
\begin{table}[!t]
\centering
\footnotesize
\setlength{\tabcolsep}{4pt}
\caption{Performance of the final \textsc{SEAM} model under the full-training regime. The model uses DistilHuBERT with 8\,s windows and top-1-layer unfreezing, trained for 3 epochs. Metrics are reported on the grouped internal \textit{eval}/\textit{test} splits and the external interview-domain set; mean and standard deviation are computed across three seeds.}
\label{tab:final}
\resizebox{\columnwidth}{!}{%
\begin{tabular}{lcccccc}
\toprule
Seed & {\bf Eval Acc} & {\bf Eval AUC} & {\bf Test Acc} & {\bf Test AUC} & {\bf Ext Acc} & {\bf Ext AUC} \\
\midrule
1337 & 0.9690 & 0.9820 & 0.9639 & 0.9781 & 0.9540 & 0.9725 \\
1338 & 0.9602 & 0.9762 & 0.9548 & 0.9715 & 0.9431 & 0.9669 \\
1339 & 0.9728 & 0.9835 & 0.9681 & 0.9802 & 0.9580 & 0.9745 \\
\midrule
\tikzmarknode{meanL}{Mean} & 0.9673 & 0.9806 & 0.9623 & 0.9766 & 0.9517 & \tikzmarknode{meanR}{0.9713} \\
\noalign{%
\begin{tikzpicture}[overlay,remember picture]
\draw[densely dotted, rounded corners, line width=0.8pt]
([xshift=-2pt,yshift=2pt]meanL.north west) rectangle
([xshift=2pt,yshift=-2pt]meanR.south east);
\end{tikzpicture}%
}
Std & 0.0065 & 0.0039 & 0.0068 & 0.0045 & 0.0077 & 0.0039 \\
\bottomrule

\end{tabular}%
}
\vspace{-0.1in}
\end{table}

\subsection{Transfer-oriented evaluation protocol}
\label{ssec:evaluation_protocol}

\skippingparagraph{\bf Internal train/eval/test protocol:}
The curated four-corpus English dataset serves as the internal development benchmark for \textsc{SEAM}. In the main training setup, we use 12 shards per class, each containing 20\,h of audio, for a total of 240\,h per class. The curated training data include 290 speakers from PodcastFillers, 1044 from People's Speech, 316 from Spoken Wikipedia, and 589 from LibriSpeech. After curation and preprocessing, these data are pooled and partitioned into grouped 80/10/10 \textit{train}/\textit{eval}/\textit{test} splits, using speaker, user, or recording-family keys where available to reduce leakage from related recordings. The \textit{eval} split is used for model selection and comparison across configurations, while the \textit{test} split is reserved for final internal reporting. All main results and ablation studies use this grouped internal protocol unless otherwise noted.

% \skippingparagraph{\bf External interview-domain evaluation:}
% Because even grouped held-out evaluation can still reward residual corpus or channel shortcuts, we add a second evaluation on a proprietary English interview dataset from the target deployment domain. This set is used for evaluation only and has no speaker overlap with the internal training data. It contains 720 clips in total (5,827 analysis windows) and deliberately crosses speaking style with channel conditions through four recording types: (i) clean scripted speech (149 clips, 1,168 windows, 34 speakers), (ii) scripted speech with mixed room and microphone conditions (187 clips, 1,542 windows, 43 speakers), (iii) clean spontaneous speech (161 clips, 1,263 windows, 38 speakers), and (iv) spontaneous speech with mixed room and microphone conditions (223 clips, 1,854 windows, 49 speakers). This construction is intentionally adversarial to shortcut-reliant models: systems that mainly exploit channel cleanliness or corpus-specific artifacts may still perform well on the internal benchmark, yet degrade substantially on this interview-domain evaluation.

\skippingparagraph{\bf External interview-domain evaluation:}
Because even grouped held-out evaluation can still reward residual corpus or channel shortcuts, we add a second evaluation on a proprietary English interview dataset from the target deployment domain. This set is used for evaluation only, has no speaker overlap with the internal training data, and is labeled independently by two human annotators, with third-annotator adjudication for disagreements. Before adjudication, annotators achieved $\kappa$ = 0.71 (84.2\% raw agreement), supporting the reliability of the clip-level labels despite mixed recording conditions. Labels are assigned and evaluated at the clip level. The benchmark contains 720 clips in total and is roughly balanced across scripted and spontaneous speech while deliberately crossing speaking style with channel conditions through four recording types: (i) clean scripted speech (149 clips, 34 speakers), (ii) scripted speech with mixed room and microphone conditions (187 clips, 43 speakers), (iii) clean spontaneous speech (161 clips, 38 speakers), and (iv) spontaneous speech with mixed room and microphone conditions (223 clips, 49 speakers). This construction is intentionally adversarial to shortcut-reliant models: systems that mainly exploit channel cleanliness or corpus-specific artifacts may still perform well on the internal benchmark, yet degrade substantially on this interview-domain evaluation.

\skippingparagraph{\bf Metrics and aggregation:}
We report window-level Accuracy and ROC-AUC for the internal \textit{eval} split, the internal \textit{test} split, and the external interview-domain set. In deployment-oriented evaluation, file-level decisions are produced by aggregating window-level outputs: we take the median of the window logits for each file, apply the sigmoid function, and then threshold the resulting score. We also tested Platt scaling~\cite{platt}, but did not use it in the final system because it introduced threshold instability across runs.

\begin{table}[!t]
\centering
\small
\caption{Effect of shortcut-prevention components under the fixed-budget regime. Uniform waveform preprocessing is kept fixed, while noise-bank augmentation and seam-aware sampling are ablated individually and jointly.}
\label{tab:shortcut_ablation}
\resizebox{\columnwidth}{!}{%
\begin{tabular}{lcccc}
\toprule
Setting & {\bf Eval AUC} & {\bf Test AUC} & {\bf Ext Acc} & {\bf Ext AUC} \\
\midrule
baseline on          & 0.9550 & 0.9287 & \bgreen{0.8527} & \bgreen{0.8991} \\
seam off             & 0.9638 & 0.9328 & 0.8179 & 0.8674 \\
specaug (seam on)    & 0.9653 & 0.9405 & 0.7807 & 0.8202 \\
noise off            & 0.9792 & 0.9491 & 0.7089 & 0.7518 \\
noise off + seam off & \bred{0.9848} & \bred{0.9557} & 0.6882 & 0.7324 \\
\bottomrule
\end{tabular}%
}
\vspace{-0.1in}
\end{table}

\vspace{-0.1in}
\section{Results}

We report results in two experimental regimes. \textbf{(i) Full-training regime:} the primary reported system is trained for three epochs on the full internal dataset (12 shards per class; 240\,h per class) and evaluated on the grouped internal \textit{eval}/\textit{test} splits and the external interview-domain set (Table~\ref{tab:final}). \textbf{(ii) Fixed-budget regime:} to compare architectural and data-design choices efficiently, we run one-epoch ablations under a reduced, fixed-budget setup (batch size 256, 4 shards per class, 120 optimization steps). Absolute metrics in this regime are not directly comparable to the full-training results; we use them to measure relative trends across design choices.

\vspace{-0.1in}
\subsection{Main results}
We first report the performance of the final \textsc{SEAM} configuration: DistilHuBERT with 8\,s windows, unfreezing the top transformer layer, and shortcut-aware training enabled (uniform preprocessing, seam-aware sampling, and non-speech noise-bank augmentation). Table~\ref{tab:final} shows mean and standard deviation across three seeds on the internal \textit{eval} split, the internal \textit{test} split, and the external interview-domain evaluation set.

Under this full-training regime, the model reaches 0.9623 $\pm$ 0.0068 test accuracy and 0.9766 $\pm$ 0.0045 test ROC-AUC on the internal test split, while also achieving 0.9517 $\pm$ 0.0077 external accuracy and 0.9713 $\pm$ 0.0039 external ROC-AUC on interview audio. The small variance across seeds indicates stable training, and the strong external performance suggests that the system is not relying primarily on corpus- or channel-specific shortcuts.

\subsection{Ablation studies}
% We next analyze the main design choices behind the final system: window length, adaptation depth, and shortcut-prevention mechanisms.

% VSEVOLOD 03/01/26 UPDATED
% \subsection{Window-length selection}
% Two-second windows miss phrase-level cues such as intonation arcs, hesitation patterns, and local timing structure, and performance drops accordingly. Performance improves substantially from 2\,s to 8\,s, with 8\,s giving the strongest held-out and interview-domain results. Moving to 12\,s remains competitive but adds latency and yields a small decline in AUC, likely because longer windows contain more silence and are more sensitive to window-to-utterance alignment.
%
% Figure~\ref{fig:window_test} shows the test curves, and Table~\ref{tab:win_ablation} summarizes the best checkpoint per window length.
\vspace{-0.1in}
\skippingparagraph{\bf Shortcut-prevention ablation:}
% Our central question is whether the shortcut-aware components improve true transfer, rather than merely acting as generic regularization. To isolate their effect, we run a one-epoch ablation at 8\,s windows with DistilHuBERT tr1 while keeping uniform preprocessing enabled and toggling only (i) the non-speech noise bank and (ii) seam-aware sampling. Table~\ref{tab:shortcut_ablation} reports results at 120 steps.
% Disabling the noise bank, and especially disabling both safeguards together, increases internal held-out AUC but sharply degrades interview-domain transfer. For example, turning off both defenses raises internal test AUC from 0.9287 to 0.9557, yet drops external AUC from 0.8991 to 0.7324. This pattern indicates that the apparent held-out gains come from reintroducing spurious corpus/channel cues rather than improving speaking-style modeling. Seam-aware sampling has a smaller effect than the noise bank on internal metrics, but it consistently improves external reliability. Overall, this ablation supports the main claim of \textsc{SEAM}: shortcut-aware data design is necessary for robust interview-domain transfer.
Table~\ref{tab:shortcut_ablation} shows that removing the shortcut-prevention components improves internal held-out metrics but reduces interview-domain transfer. Particularly, disabling both the noise bank and seam-aware sampling increases internal test ROC-AUC from 0.9287 to 0.9557 while reducing external ROC-AUC from 0.8991 to 0.7324. This supports the main claim of SEAM: shortcut-aware design improves transfer even when naive held-out performance increases without it.

\vspace{-0.05in}
\skippingparagraph{\bf Window-length selection:}
Window length materially affects performance (Table~\ref{tab:win_ablation}). Results improve from 2\,s to 8\,s, while 12\,s remains competitive but slightly worse and incurs higher latency. We therefore use 8\,s windows in the final system as the best quality--latency tradeoff.

% \begin{figure*}[t]
%   \centering
%   \begin{minipage}[t]{0.495\textwidth}
%     \centering
%     \includegraphics[width=\linewidth]{fig_window_test_acc.png}
%   \end{minipage}\hfill
%   \begin{minipage}[t]{0.495\textwidth}
%     \centering
%     \includegraphics[width=\linewidth]{fig_window_test_auc.png}
%   \end{minipage}
%   \caption{Window-length sweep: test Accuracy (left) and test ROC-AUC (right) versus training step.}
%   \label{fig:window_test}
% \end{figure*}

\begin{table}[!t]
\centering
\small
\caption{Effect of analysis window length under the fixed-budget regime.}
\label{tab:win_ablation}
\resizebox{\columnwidth}{!}{%
\begin{tabular}{ccccc}
\toprule
Win(s) & {\bf Test Acc} & {\bf Test AUC} & {\bf Ext Acc} & {\bf Ext AUC} \\
\midrule
2  & 0.7229 & 0.7739 & 0.6572 & 0.7485 \\
4  & 0.8629 & 0.8889 & 0.8067 & 0.8646 \\
8  & \textbf{0.8949} & \textbf{0.9287} & \textbf{0.8527} & \textbf{0.8991} \\
12 & 0.8786 & 0.9067 & 0.8349 & 0.8813 \\
\bottomrule
\end{tabular}%
}
\vspace{-0.05in}
\end{table}

% VSEVOLOD 03/01/26 UPDATED
% \subsection{How much to unfreeze?}
% Table~\ref{tab:unfreeze_ablation} shows that unfreezing only the top transformer layer gives the best out-of-domain results. Unfreezing more layers improves adaptation capacity, but that extra flexibility does not translate into better cross-corpus generalization here. Best checkpoints are selected by eval AUC within each setting.

% \skippingparagraph{\bf How much to unfreeze?}
% Table~\ref{tab:unfreeze_ablation} shows that shallow partial fine-tuning is preferable to both weaker and stronger adaptation. Training only the classifier head underfits, while unfreezing the top transformer layer yields the strongest performance. Unfreezing additional layers increases adaptation capacity but does not improve generalization in this setting; the top-2 and CNN-inclusive variants trail the top-1 configuration. We therefore use top-1-layer unfreezing in the final model.
\vspace{-0.05in}
\skippingparagraph{\bf How much to unfreeze?}
Table~\ref{tab:unfreeze_ablation} shows that shallow partial fine-tuning is preferable to both weaker and stronger adaptation. Training only the classifier head underperforms, and unfreezing the top transformer layer gives the best results, so we use top-1-layer unfreezing in the final model.

\begin{table}[!t]
\centering
\small
\caption{Effect of adaptation depth under the fixed-budget.}
\label{tab:unfreeze_ablation}
\begin{tabular*}{\columnwidth}{@{\extracolsep{\fill}}lccc}
\toprule
Setting & {\bf Eval AUC} & {\bf Test Acc} & {\bf Test AUC} \\
\midrule
head & 0.7799 & 0.7078 & 0.7625 \\
tr1  & \textbf{0.9550} & \textbf{0.8949} & \textbf{0.9287} \\
tr2  & 0.9296 & 0.8502 & 0.9031 \\
cnn2 & 0.7631 & 0.6887 & 0.7354 \\
\bottomrule
\end{tabular*}
\vspace{-0.05in}
\end{table}

\begin{table}[!t]
\centering
\scriptsize
\setlength{\tabcolsep}{4pt}
\caption{Post-training quantization results for the final model on NVIDIA L4, evaluated on the external interview-domain set. Quantization substantially reduces memory footprint with little change in external accuracy or ROC-AUC.}
\label{tab:quant}
\resizebox{\columnwidth}{!}{%
\begin{tabular}{lccccc}
\toprule
Precision & \textbf{VRAM (MB)} & \textbf{Ext Acc (\%)} & \textbf{Ext AUC} & \textbf{Full Pipe (ms/win)} \\
\midrule
AMP (base) & 90.37 & 95.17 & 0.9713 & 6.99 $\pm$ 0.010 \\
INT8       & 48.74 & 95.30 & 0.9700 & 7.86 $\pm$ 0.005 \\
INT4       & 41.80 & 95.35 & 0.9743 & 7.25 $\pm$ 0.009 \\
\bottomrule
\end{tabular}%
}
\vspace{-0.05in}
\end{table}

\begin{table}[t]
\centering
\small
\caption{Backbone screening (frozen encoder with MLP head, fixed-budget) and streaming inference efficiency on NVIDIA L4 (12\,s windows, batch size 1).}
\label{tab:backbone_screening_eff}
\resizebox{\columnwidth}{!}{%
\begin{tabular}{lccrrrr}
\toprule
\multirow{2}{*}{\textbf{Backbone}} &
\multicolumn{2}{c}{\textbf{Test Set}} &
\multicolumn{4}{c}{\textbf{Streaming efficiency}} \\
\cmidrule(lr){2-3}\cmidrule(lr){4-7}
& \textbf{Acc} & \textbf{AUC} & \textbf{Params(M)} & \textbf{MB} & \textbf{RTF} \\
\midrule
DistilHuBERT      & 0.754 & 0.848 & 23.49 &  93.97 & 5.60e-4 \\
Distil-wav2vec2   & 0.736 & 0.822 & 51.84 & 207.52  & 6.76e-4 \\
HuBERT Base       & 0.779 & 0.868 & 94.37 & 377.57  & 8.37e-4 \\
wav2vec2 Base     & 0.778 & 0.868 & 94.37 & 377.61 & 8.33e-4 \\
WavLM Base+       & \textbf{0.803} & \textbf{0.895} & 94.38 & 377.62 & 1.53e-3 \\
\bottomrule
\end{tabular}%
}
\end{table}

% \skippingparagraph{\bf Backbone selection and deployment tradeoff:}
% Table~\ref{tab:backbone_screening} summarizes ROC-AUC for candidate backbones trained with an MLP head only under a fixed screening setup, while Table~\ref{tab:eff} shows the corresponding deployment costs. WavLM Base+ achieves the highest ROC-AUC (0.895), with DistilHuBERT lower at 0.849. However, DistilHuBERT is substantially lighter and faster, with 23.49\,M parameters, a 93.97\,MB checkpoint, 6.72\,ms latency per 12\,s window, and a real-time factor of $5.60\times10^{-4}$ on an NVIDIA L4. Compared with larger encoders such as HuBERT Base, wav2vec2 Base, and WavLM Base+, it offers a much smaller deployment footprint while remaining competitive after task-specific tuning. We therefore select DistilHuBERT for the final system because it lies on the most favorable quality--efficiency frontier for real-time scripted-versus-spontaneous monitoring.
\vspace{-0.05in}
\skippingparagraph{\bf Backbone selection and deployment tradeoff:}
As shown in Table~\ref{tab:backbone_screening_eff}, WavLM Base+ gives the highest screening ROC-AUC, but DistilHuBERT offers a substantially smaller and faster deployment footprint while remaining competitive. We therefore select DistilHuBERT as the most favorable quality--efficiency tradeoff for real-time use.

\begin{figure}[t]
  \centering
  \includegraphics[width=\columnwidth]{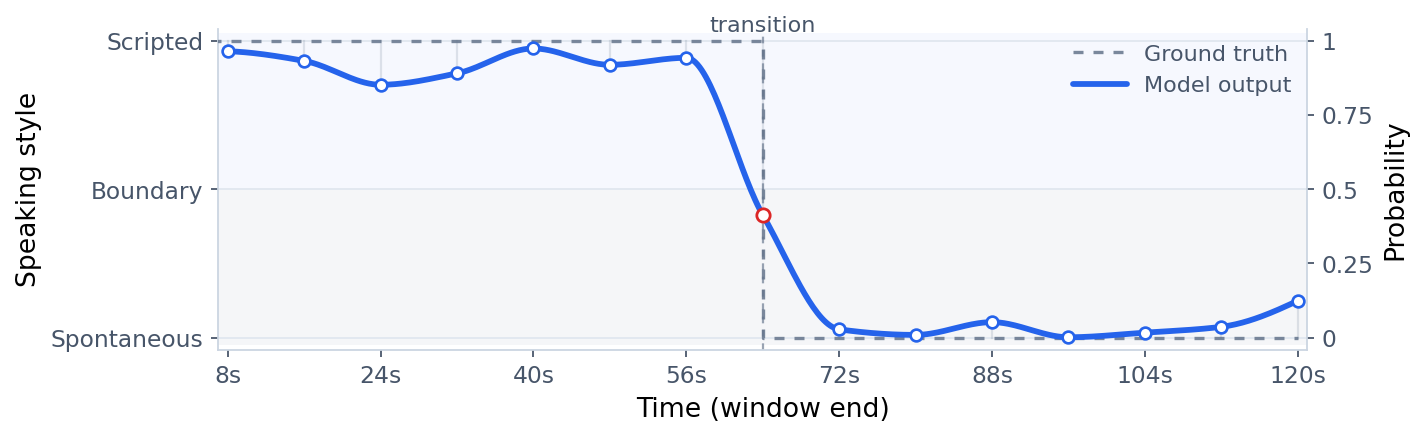}
  \caption{Window-level scriptedness scores for a natural within-speaker transition over 120\,s. Predictions are computed on 8\,s windows; the solid curve shows the model output and dashed step function indicates corresponding window labels.}
  \label{fig:transition_audit}
\end{figure}

\vspace{-0.05in}
\skippingparagraph{\bf Quantized deployment variants:}
Table~\ref{tab:quant} shows that post-training quantization reduces footprint from 90.37\,MB to 48.74\,MB (INT8) and 41.80\,MB (INT4) with little change in external accuracy or ROC-AUC. This indicates that the final model can be compressed aggressively while preserving deployment-facing robustness.
\vspace{-0.05in}
\skippingparagraph{\bf Qualitative transition audit:}
Figure~\ref{fig:transition_audit} shows a natural within-speaker shift from more scripted to more spontaneous delivery over 120\,s. The predicted score changes near the transition and remains stable before and after it, consistent with tracking speaking style rather than obvious edit artifacts.

\vspace{-0.5in}
\section{Discussion and Limitations}
Our results suggest that robust scripted-versus-spontaneous detection depends not only on the backbone, but on shortcut-aware data design and evaluation. In particular, disabling the shortcut-prevention components can improve held-out performance while degrading interview-domain transfer, indicating that standard held-out results may still reward corpus- or channel-specific heuristics than genuine speaking-style cues.

At the same time, these methods do not remove dataset fingerprints completely. Scriptedness remains entangled with genre, recording practice, and corpus construction, so the current system should be viewed as a narrow guardrail signal rather than a complete representation of speaking style. Our setup is also English-first and we do not make a multilingual claim. As a robustness check, we evaluate zero-shot transfer to five held-out non-English languages. Performance remains above chance overall, suggesting that the model captures some language-agnostic cues, but the drop relative to English and the variation across languages show that this transfer is only partial.

Finally, the interview-domain benchmark is deliberately challenging, but still limited relative to real deployment. Future work should expand evaluation across broader languages, accents, devices, and environments, and improve calibration under distribution shift.

\vspace{-0.1in}
\section{Conclusion and Future Work}
We presented \textsc{SEAM}, a compact real-time framework for scripted-versus-spontaneous speech detection that emphasizes shortcut-aware training and evaluation. Results show that improving robustness to corpus and channel shortcuts is important for transfer to interview-domain audio, while still supporting practical low-latency deployment. Future work will extend evaluation across broader languages and recording conditions, and improve calibration and temporal aggregation under distribution shift.

% \section{Acknowledgments}
% We thank colleagues at Symbal AI and Washington State University for discussions on deployment constraints and evaluation.
\vspace{-0.1in}
\section{Generative AI Use Disclosure}
Generative AI tools were used for language polishing and LaTeX formatting assistance. All authors reviewed, edited, and verified the final content, results, and claims.

\FloatBarrier

%===============================================================================
% References (IEEE style)

\end{document}